\documentclass[aps,prb,twocolumn,preprintnumbers,amsmath,amssymb,superscriptaddress]{revtex4}
\pdfoutput=1
\usepackage{graphicx}
\usepackage{textcomp} 
\usepackage[utf8]{inputenc}
\usepackage{xcolor}
\usepackage[final]{pdfpages}
\usepackage{miller}
\usepackage{changes}

\usepackage[colorlinks=true, linkcolor=blue, citecolor=blue, filecolor=blue, urlcolor=blue,pdfproducer={.},pdfcreator={.}]{hyperref}
\usepackage{cleveref}

\begin{document}

\newcommand{\gpi}{\textrm{\greektext p}}
\newcommand{\gmu}{\textrm{\greektext m}}

\newcommand{\highl}[1]{{\color{red}#1}}
\newcommand{\commnt}[1]{{\it \color{blue}#1}}

\newcommand{\zit}[1]{[#1]}

    \title{Indirect optical manipulation of the antiferromagnetic order of insulating NiO\\ by ultrafast interfacial energy transfer}
    
    \author{Stephan~\surname{Wust}}
    \email{wust@rhrk.uni-kl.de}
    	\affiliation{Department of Physics and Research Center OPTIMAS, RPTU University Kaiserslautern-Landau, 67663 Kaiserslautern, Germany}
    
    \author{Christopher~\surname{Seibel}}
    	\affiliation{Department of Physics and Research Center OPTIMAS, RPTU University Kaiserslautern-Landau, 67663 Kaiserslautern, Germany}
    
    \author{Hendrik~\surname{Meer}}
    	\affiliation{Institute of Physics, Johannes Gutenberg-University Mainz, 55099 Mainz, Germany}
    
    \author{Paul~\surname{Herrgen}}
    	\affiliation{Department of Physics and Research Center OPTIMAS, RPTU University Kaiserslautern-Landau, 67663 Kaiserslautern, Germany}
    
    \author{Christin~\surname{Schmitt}}
    	\affiliation{Institute of Physics, Johannes Gutenberg-University Mainz, 55099 Mainz, Germany}
    
    \author{Lorenzo~\surname{Baldrati}}
    	\affiliation{Institute of Physics, Johannes Gutenberg-University Mainz, 55099 Mainz, Germany}
    
    \author{Rafael~\surname{Ramos}}
    	\affiliation{CIQUS, Departamento de Qu\'imica-F\'isica, Universidade de Santiago de Compostela, Santiago de Compostela, Spain}
    	\affiliation{WPI Advanced Institute for Materials Research, Tohoku University, Sendai 980-8577, Japan}
    
	\author{Takashi Kikkawa}
		\affiliation{Department of Applied Physics, The University of Tokyo, Tokyo 113-8656, Japan}
        \affiliation{Advanced Science Research Center, Japan Atomic Energy Agency, Tokai 319-1195, Japan}
    
    \author{Eiji Saitoh}
		\affiliation{Department of Applied Physics, The University of Tokyo, Tokyo 113-8656, Japan}
		\affiliation{Institute for AI and Beyond, The University of Tokyo, Tokyo 113-8656, Japan}
		\affiliation{RIKEN Center for Emergent Matter Science (CEMS), Wako 351-0198, Japan}
		\affiliation{WPI Advanced Institute for Materials Research, Tohoku University, Sendai 980-8577, Japan}
    
	\author{Olena~\surname{Gomonay}}
		\affiliation{Institute of Physics, Johannes Gutenberg-University Mainz, 55099 Mainz, Germany}
		
	\author{Yuriy~\surname{Mokrousov}}
		\affiliation{Institute of Physics, Johannes Gutenberg-University Mainz, 55099 Mainz, Germany}
		\affiliation{Peter Gr\"unberg Institut and Institute for Advanced Simulation, Forschungszentrum J\"ulich and JARA, 52425 J\"ulich, Germany.}
		
    \author{Hans~Christian~\surname{Schneider}}
    	\affiliation{Department of Physics and Research Center OPTIMAS, RPTU University Kaiserslautern-Landau, 67663 Kaiserslautern, Germany}
    	
    \author{Mathias~\surname{Kl\"aui}}
    	\affiliation{Institute of Physics, Johannes Gutenberg-University Mainz, 55099 Mainz, Germany}
    	
    \author{Baerbel~\surname{Rethfeld}}
    	\affiliation{Department of Physics and Research Center OPTIMAS, RPTU University Kaiserslautern-Landau, 67663 Kaiserslautern, Germany}
    	
    \author{Benjamin~\surname{Stadtm\"uller}}
    	\email{benjamin.stadtmueller@uni-a.de}
    	\affiliation{Department of Physics and Research Center OPTIMAS, RPTU University Kaiserslautern-Landau, 67663 Kaiserslautern, Germany}
    	\affiliation{Experimental Physics II, Institute of Physics, University of Augsburg, Germany}
    \author{Martin~\surname{Aeschlimann}}
    	\affiliation{Department of Physics and Research Center OPTIMAS, RPTU University Kaiserslautern-Landau, 67663 Kaiserslautern, Germany}
    \date{\today}

\begin{abstract}
We report the ultrafast, (sub)picosecond reduction of the antiferromagnetic order in an insulating NiO thin film within a Pt/NiO bilayer structure following optical excitation with near-infrared femtosecond pulses. Unlike in pure NiO films, where no significant magnetic response is observed under similar excitation conditions, the Pt/NiO bilayer exhibits a pronounced, non-equilibrium decrease in its magneto-optical birefringence signal on the sub-picosecond timescale. This effect is attributed to an efficient interfacial energy transfer from the optically excited electron system of the Pt layer into the spin system of NiO, as
supported by temperature-based 
model simulations
which reproduce the observed dynamics. 
We propose that this interfacial energy transfer process is accompanied by an interfacial exchange-mediated angular momentum transfer from the spin angular momentum of Pt hot electrons to the spin of NiO. The accumulation of spin angular momentum in Pt hot electrons is attributed to spin-orbit coupling in Pt and the magnetic proximity effect at the Pt/NiO interface. 
 Our results reveal a general pathway for the indirect, ultrafast manipulation of the magnetic order in insulating and semiconducting antiferromagnets through optimized interfacial engineering, opening new possibilities for all-optical control in antiferromagnetic spintronics.
\end{abstract}

\maketitle

\section{Introduction}
One of the most crucial challenges in antiferromagnetic spintronics is to optimize the speed and efficiency of the manipulation of the antiferromagnetic  order that is characterized by the N\'eel vector. While current-driven manipulation of the N\'eel order by spin-orbit torques is a highly promising and reliable approach for storing information in antiferromagnets (AFMs) \cite{bodnar2018writing, olejnik2018terahertz, meer2020}, its operation speed is still limited to the timescale of current pulses, which is typically in the order of several tens to hundreds of picoseconds \cite{vzelezny2018spin, jungwirth2018multiple}. Faster manipulation of the N\'eel order can be achieved by optical excitation of the AFM spin system.  Pioneering studies have demonstrated the collective excitation of the spin system by THz and near-infrared (near-IR) excitation \cite{olejnik2018terahertz, qiu2021ultrafast, tzschaschel2017ultrafast, bossini2021ultrafast, Dabrowski2026}. Due to the electrically insulating nature of many AFMs, the spin system is directly excited by the electromagnetic light field via resonant excitation by the magnetic field component of the radiation \cite{kampfrath2011coherent}, inelastic impulsive stimulated Raman scattering \cite{bossini2016macrospin}, or d-d excitations \cite{wang2022ultrafast, gillmeister2020ultrafast}. All these excitations result in coherent precessions of the N\'eel vector with frequencies of up to a few THz corresponding to a response of the spin-system on the picosecond timescale \cite{bossini2021ultrafast, satoh2017excitation}. 

An even faster response of the spin system of AFMs can potentially be achieved by strong non-equilibrium excitation of the material using fs-light pulses. This approach has already been demonstrated extensively for ferromagnetic materials for which an optical excitation with near-IR fs light pulses results in an incoherent loss of the magnetic order within a few hundreds of femtoseconds \cite{Beaurepaire1996, koopmans2005unifying, medapalli2012efficiency, battiato2010superdiffusive, you2018revealing, Zhang2000laserinduced, bigot2005ultrafast}. 
This so-called ultrafast demagnetization is however predominantly observed for metallic (ferromagnetic) systems for which the optical excitation results in a strong non-equilibrium excitation of the material's electron system, which subsequently transfers its energy to the spin system 
\cite{koopmans2010explaining,2013feedback}
. This excitation scheme is, however, not possible for a wide range of AFMs, which are largely insulators with bandgaps of several eV. 

In our work, we demonstrate an alternative approach to incoherently manipulate the spin order of insulating antiferromagnet nickel oxide (NiO) on (sub-)picosecond timescale using fs near-IR pulses with photon energies below the bandgap of the insulating material. This is achieved by an ultrafast energy transfer from an adjacent metallic platinum (Pt) layer that is strongly excited by the ultrashort near-IR pulse. We propose that this ultrafast energy transfer occurs directly between the hot electron system of the metallic layer and the spin system of the insulating antiferromagnet and is mediated by a stochastic exchange scattering at the interface.

\section{Results and Discussion}

We start with our experimental investigation of the ultrafast response of NiO.
NiO is a collinear wide-bandgap AFM ($E_{\rm gap}=4.3\,$eV~\cite{sawatzky1984}) and a promising candidate for possible spintronic applications due to its high N\'eel temperature of $T_{N}=523\,$K in the bulk \citep{Roth1960}, the possibility to electrically read and manipulate the antiferromagnetic order \citep{Hoogeboom2017, Moriyama2018, Schmitt2020b}, and the observations of ultrafast currents in the THz regime in Pt/NiO bilayers structures \cite{kampfrath2011coherent, Moriyama2020}. Below $T_{N}$, NiO is an easy plane antiferromagnet with a ferromagnetic exchange coupling of the spins within the $\hkl{111}$ planes and an antiferromagnetic coupling between spins of adjacent  $\hkl{111}$ planes. Due to the NiO crystal symmetry, this results in the formation of four differently oriented twin-domains (T-domains), each consisting of three spin-domains (S-domains). In a bulk crystal, this leads to the coexistence of 12 antiferromagnetic domains \citep{Roth1960, Roth1960a, Slack1960}. 

In our study, we focus on a $10\,$nm thick NiO epitaxial thin film capped with a $2\,$nm Pt layer grown on a MgO(001) substrate \citep{Schreiber2020}. In this case, the additional strain from the substrate leads to a preferential out-of-plane orientation of the N\'eel vector of NiO \citep{ Alders1998, Altieri2003}. This reduces the number of S-domains for each T-domain and in essence, leads to a single S-domain state within each T-domain \cite{Schmitt2020b}.

\begin{figure}[t]
    \centering
    \includegraphics{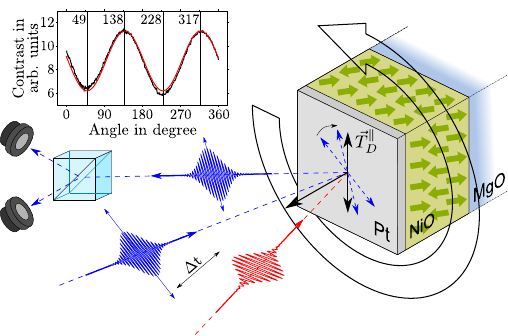}
    \caption{Schematic illustration of the experimental approach to record the magnetic response of NiO by is magneto-optical birefringence (MOBF) signal. Changing the azimuthal orientation between the polarization of the probe beam and the projected in-plane anisotropy axis $\vec{T}_{D}^{\parallel}$ of the AFM NiO results in a significant change in the rotation of the light polarization as shown in the inset. The response of the MOBF signal after a time $\Delta$t after the optical excitation is monitored in a pump-probe experiment. See supporting material for more details on the experiment.}
    \label{fig:exp_setup}
\end{figure}

The ultrafast response of the spin system of NiO is monitored experimentally by the change in the magneto-optical birefringence (MOBF) signal in an all-optical pump-probe setup as shown in Fig.~\ref{fig:exp_setup}. The MOBF of NiO is the result of a rhombohedral distortion of the NiO crystal lattice in its AFM phase which was observed both for bulk crystals \citep{Germann1974} as well as for NiO thin films \citep{Xu2019a, Schreiber2020}.  This rhombohedral distortion associated with the magnetic AFM order leads to an anisotropy axis along the direction of the T-Domain $\vec{T}_{D}$ ($\hkl<111>$) and thus to a different reflectivity of light with parallel and perpendicular polarization with respect to this anisotropy axis. This difference in reflectivity of the light field components can be detected as a rotation of light polarization of linearly polarized laser light after the reflection from the sample. 
We monitor the change in the orientation of the linearly polarized probe laser beam (photon energy $3.1\,$eV) in our experiment by using a polarization-sensitive balancing detector \cite{Xu2019a}, see Fig.~\ref{fig:exp_setup}. In normal incidence geometry, we can continuously rotate the azimuthal orientation of the light polarization within the surface plane, which results in a sinusoidal oscillation of the detected signal (further explained in supplementary material). The periodicity of $180^\circ$ of the signal matches the symmetry of the rhombohedral distortion of the NiO crystal lattice. This clearly demonstrates that our normal incidence experiment is sensitive enough to detect the magneto-optical birefringence signal despite the fact that we are only able to access the projection of the anisotropy axis in the plane of the sample surface $\vec{T}_{D}^{\parallel}$.

In our pump-probe experiment, we use an almost collinear normal incidence geometry of the linearly polarized pump and probe beam. The photon energy of the pump beam is $1.55\,$eV and hence significantly smaller than the optical band gap of NiO. We recorded time-resolved MOBF traces for two orientations of $45^\circ$ and $-45^\circ$ of the probe beam polarization with respect to the anisotropy axis. Subtracting these traces with antisymmetric MOBF contributions (see inset Fig.~\ref{fig:exp_setup}) from each other allows us to eliminate pure optical signals that are expected to be symmetric in both traces.

\begin{figure}
    \centering
    \includegraphics[width=\linewidth]{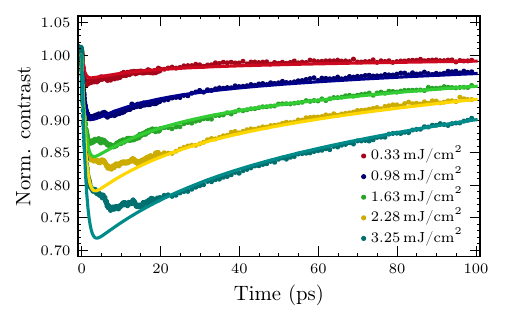}
    \caption{Temporal evolution of the MOBF signal for the Pt($2\,$nm)/NiO($10\,$nm) bilayer structure after the optical excitation with $1.55\,$eV pump photons, and normal incidence geometry. Experimental data are shown as colored dots, while the solid lines represent simulated MOBF traces that were obtained by a temperature-based model simulation.}
    \label{fig:mo_ptnio}
\end{figure}

\begin{figure}
    \centering
    \includegraphics[width=\linewidth]{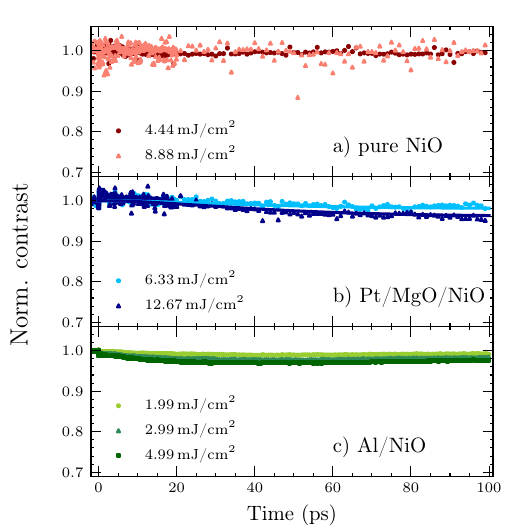}
    \caption{Comparison of the temporal evolution of the MOBF signal (a) pure NiO, (b) a $3\,$nm MgO spacer between Pt and NiO, and (c) a Al($4\,$nm)/NiO($10\,$nm) multilayer structure. For a better comparison, the same scale as in Fig.~\ref{fig:mo_ptnio} is used. Next to experimental data shown as colored dots in all panels, (b) shows solid lines representing simulated MOBF traces obtained from a temperature-based model.}
    \label{fig:mo_comparison}
\end{figure}

We start the discussion of our experimental findings with the ultrafast magnetic response of the Pt/NiO(001) bilayer system, shown in Fig.~\ref{fig:mo_ptnio}. After optical excitation in the near-IR range ($\hbar\omega=1.55\,$eV), the MOBF signal reduces significantly on a fs timescale followed by a slow relaxation of the signal on a timescale of several tens of picoseconds. The reduction of the MOBF signal is attributed to a reduction of the antiferromagnetic order in NiO. This conclusion is based on a recent time-resolved diffraction study which revealed a clear correlation between the reduction of the rhombohedral distortion of NiO and the change in the AFM order of the system~\cite{windsor2021exchange}.
Moreover, the transient MOBF traces reveal a coherent oscillation which will later be adressed in more detail.

This situation changes drastically for the bare NiO(001) thin film, shown in Fig.~\ref{fig:mo_comparison}(a). 
After excitation with the same photon energy as for the Pt/NiO bilayer system, we find no change in the transient MOBF signal for the bare NiO thin film. This is not surprising when considering NiO's large bandgap of $4.3\,$eV, which would require a three-photon absorption process 
or UV excitation \cite{gillmeister2020ultrafast}
to optically excite the NiO electron system. Such a process is, however, extremely unlikely for the absorbed laser fluences used in our experiment \cite{sun2017high}. 

Overall, our experimental data clearly demonstrate an optically induced loss of the antiferromagnetic order for the Pt/NiO bilayer system after excitation in the near-IR range that is absent for the bare NiO film for comparable fluences. In addition, the almost linear correlation between the applied fluence and the reduction of the magnetic order parameter indicates that the ultrafast loss of antiferromagnetic order is due to a linear absorption of light in the material stack. All these experimental observations point to light absorption in the metallic Pt layer and a subsequent ultrafast energy transfer from the Pt into the NiO layer. 

To further examine the microscopic origin of the indirect excitation of the NiO spin system via the Pt layer, we repeated the same experiment for a similar structure with an additional $3\,$nm thick MgO buffer layer between Pt and the NiO layers. The introduction of an insulating interlayer suppresses any interaction between the electron system of Pt and the electron or spin system of NiO, respectively \cite{mihalceanu2017spin}. The experimental MOBF traces for the Pt/MgO/NiO multilayer structure are shown in Fig.~\ref{fig:mo_comparison}(b). They only reveal a marginal reduction of the MOBF signal even for considerably larger fluences in Pt that evolves on the timescale of several tens of ps. This rather slow and inefficient coupling between the Pt and NiO layer via MgO can only be explained by a phonon-mediated energy transfer. Hence, this reference experiment also points to an efficient coupling between the Pt electron systems with the NiO spin system in the Pt/NiO bilayer structure. 

In a third reference experiment, the Pt capping layer was replaced by a 4\,nm-thick Al layer. The corresponding measurements are shown in Fig.~\ref{fig:mo_comparison}(c). Similar to the Pt/MgO/NiO sample, the antiferromagnetic order parameter decreases on a significantly longer timescale than in the Pt/NiO bilayer. This demonstrates that interfacial energy transfer from the Al layer is substantially less efficient and slower than that from the Pt overlayer. The dynamics are instead largely consistent with electron-phonon-mediated energy transfer, as observed in the Pt/MgO/NiO sample. These results indicate that the ultrafast interfacial effect in the Pt/NiO bilayer cannot be attributed solely to the presence of photoexcited hot electrons at the interface, but rather to the strong atomic spin-orbit coupling of Pt as well as the magnetic proximity effect at interfaces between Pt and magnets.

\begin{figure}[t]
    \centering
    \includegraphics[width=\linewidth]{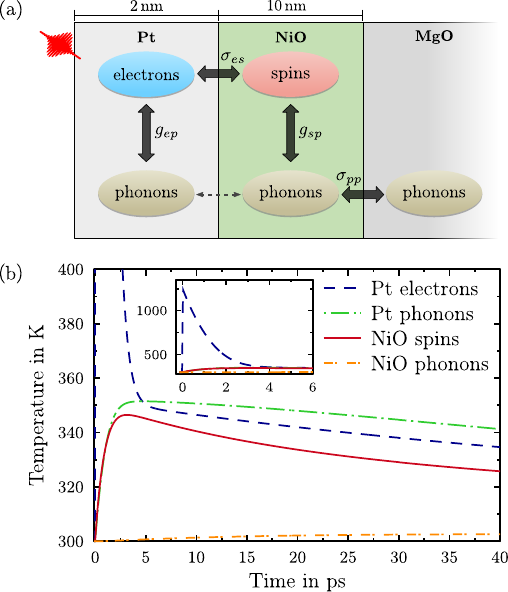}
    \caption{(a) Illustration of the electron, spin and phonon subsystems of the different layers of the Pt/NiO bilayer system on a MgO substrate that are considered in our TTM simulations. The interaction channels are marked by thick black arrows between the subsystems. (b) Temporal evolution of the simulated temperatures in Pt and NiO. The temperature of the NiO spins determines the MOBF contrast.}
    \label{fig:model}
\end{figure}

Next, we turn to our temperature-based phenomenological simulations which allow us to gain insights into the energy transfer processes within the magnet bilayer system.
Figure~\ref{fig:model}(a) illustrates the different degrees of freedom considered in our model as well as their coupling parameters. Similar to the conventional two-temperature model (TTM)~\cite{Anisimov1974}, we describe the dynamics of the metallic Pt layer by electrons coupled to phonons via an electron-phonon coupling parameter $g_{ep}$.
The heat capacity of the electrons is dynamically calculated from a realistic density of states (DOS) obtained from density functional theory (DFT)~\cite{Lin2008}. The heat capacity of the phonon system is kept constant according to the Dulong-Petit law. 

\textcolor{black}{For the NiO layer,} we can neglect the electron system due to the small photon energy of the optical excitation compared to the bandgap. However, we treat the spin system as a separate subsystem as it is frequently done when modeling magnetization dynamics on ultrafast timescales \cite{koopmans2010explaining, Beaurepaire1996}.
The heat capacity of the spin system is obtained from the measured total heat capacity~\cite{Keem1978}, subtracting the phononic contribution. It exhibits a sharp peak at the N\'eel temperature due to the antiferromagnetic/paramagnetic phase transition~\cite{Radwanski2008}.
The coupling between the spin system and the phonons is mediated by the magneto-elastic (spin-phonon) coupling $g_{sp}$~\cite{Wall2009, Aytan2017}.

The most important parameters in our simulation describe the coupling mechanisms across the Pt/NiO interface. 
The large bandgap of NiO prevents particle (charge carrier) transport across the interface. 
In contrast, energy can be exchanged between both layers by several processes. 
The laser-heated electron system of Pt can couple directly to 
the phonons of the insulating NiO \cite{huberman1994electronic,
hopkins2009effects,sokolowski2015thickness,Lombard2014}.
However, this coupling
was found to be important only on the ps to ns timescales~\cite{Lombard2014}, \textcolor{black}{as can be seen in Fig.~\ref{fig:mo_comparison}b and ~\ref{fig:mo_comparison}c, }
and is therefore not considered in our model.
Instead, the main interaction channel
involves a direct interaction of the Pt electrons with the NiO spins described by the electron-spin coupling parameter $\sigma_{es}$. 

The coupling between the Pt and NiO subsystems is calculated by solving the  
differential equations of the 
coupled layer-TTMs 
numerically~\cite{CrankNicolson1947}. The full set of equations is discussed in the supporting material. The energy or heat is distributed homogeneously within the different subsystems of the materials and we do not explicitly consider any depth-dependent variations within a single layer. 

The optical excitation is modeled by a Gaussian laser pulse of varying fluence and a pulse duration of $50\,$fs (full width at half maximum) which excites the electron system of Pt.  
Figure~\ref{fig:model}(b) shows the results of our simulation for an exemplary laser fluence, and a set of typical coupling parameters\textcolor{black}{, see supplemental material}. 
It thus shows
the time-dependent temperature changes of all subsystems after optical excitation.
The optically excited Pt electron system transfers its energy quickly to the Pt phonon system as well as into the NiO spin system thereby increasing the temperatures of these two systems. 
This results in equilibration of these three sub-systems within the first $5\,$ps to almost the same quasi-equilibrium temperature~\cite{Pudell2018}. 
Subsequently, all three subsystems reduce their temperature by dissipating energy into the still cold NiO phonon system.

To compare our simulations with the experimental findings, we translate the spin temperature into the experimental signal $\delta_\text{MOBF}$. 
We assume a quasi-instantaneous correlation between the lattice distortion and the temperature-induced changes of the antiferromagnetic order parameter~\cite{windsor2021exchange} 
and exploit the known temperature-dependence of the MOBF, which is given as~\cite{Saidl2017} 

\begin{equation}
\label{eqn:eq1}
\delta_\text{MOBF} \propto \langle M\rangle_T^2 \propto (T_N - T_S)^{2\beta}\enspace,
\end{equation}
%\cite{Saidl2017}, 
where $M$ is the magnetization of the sublattices, $T_S$ is the simulated temperature of the NiO spin system, $T_N = 523\,$K is the N\'eel temperature of NiO~\cite{Xu2019a}, and $\beta = 0.367$ is a critical exponent~\cite{Saidl2017}. 
We compare the resulting theoretical MOBF trace with the experimental curve. 
We choose the data obtained for a fluence of $1.63\,$mJ cm$^2$ to optimize our four coupling parameters $g_{ep}, \sigma_{es}, g_{sp}$ and $\sigma_{pp}$, treated as independent on temperature.
Here, $g_{ep}$ and $\sigma_{es}$  mainly drive the depth and time of the signal's minimum, respectively; $g_{sp}$ the recovery-rate of the signal within tens of ps, and $\sigma_{pp}$ the long-time behaviour towards the $100$\,ps-scale.
Then, we calculate theoretical MOBF traces for the other fluences applied in the experiment. 

The resulting curves show a very good fit with the experimental data, see 
Fig.~\ref{fig:mo_ptnio}. 
Our simulations reproduce the general trend observed in the experiment, i.e., we find a fast, (sub-) picosecond reduction of the magnetic signal followed by a significantly slower \textcolor{black}{recovery} of the magnetic order. 
The MOBF dynamics observed for the Pt/NiO bilayer stack are very similar to the so-called demagnetization curves of metallic ferromagnetic materials after strong, non-equilibrium excitation in the near-IR range~\cite{Beaurepaire1996, koopmans2005unifying, medapalli2012efficiency, battiato2010superdiffusive, you2018revealing, Zhang2000laserinduced, bigot2005ultrafast, 2013feedback}. 
However, in our case, the time constant of the loss of the antiferromagnetic order is between $0.5$ and $1.0\,$ps and hence significantly longer than for a thin metallic ferromagnetic film, such as Ni or Co on an insulating substrate~\cite{Beaurepaire1996, bigot2005ultrafast}. 
\textcolor{black}{{The timescale of the loss of antiferromagnetic order in NiO }can be attributed to the indirect heating through the optical excitation of the adjacent non-magnetic metal layer. \textcolor{black}{It} is directly reflected in the temperature evolution of the NiO spin system shown in Fig.~\ref{fig:model}(b), which is heated by the Pt electrons within 500$\,$fs to 1$\,$ps.}
The ultrafast energy transfer is thus mediated by the direct coupling between the Pt electron and NiO spin system at the interface. 
\textcolor{black}{The slower timescale of recovery of the MOBF signal in Fig.~\ref{fig:mo_ptnio} can be attributed to the characteristic time of energy transfer between the NiO spin system and phonons.}

\textcolor{black}{
Introducing an insulating layer between Pt and NiO can suppress the ultrafast interaction channel at the interface between these two materials. This is the situation studied in the reference experiment, see Fig.~\ref{fig:mo_comparison}(b). 
Accordingly, we can also suppress this channel in the simulation by setting the parameter $\sigma_{es} = 0$.} 
We show the results of the corresponding simulation for the Pt/MgO/NiO multilayer system by a solid curve superimposed onto the experimental data in Fig.~\ref{fig:mo_comparison}(b). 
\textcolor{black}{The good comparison confirms that the} 
excitation of the spin system of NiO 
through phonon-phonon coupling  
is considerably slower than the direct electron-spin coupling of the Pt/NiO system. 

\begin{figure}[t]
    \centering
    \includegraphics[width=\linewidth]{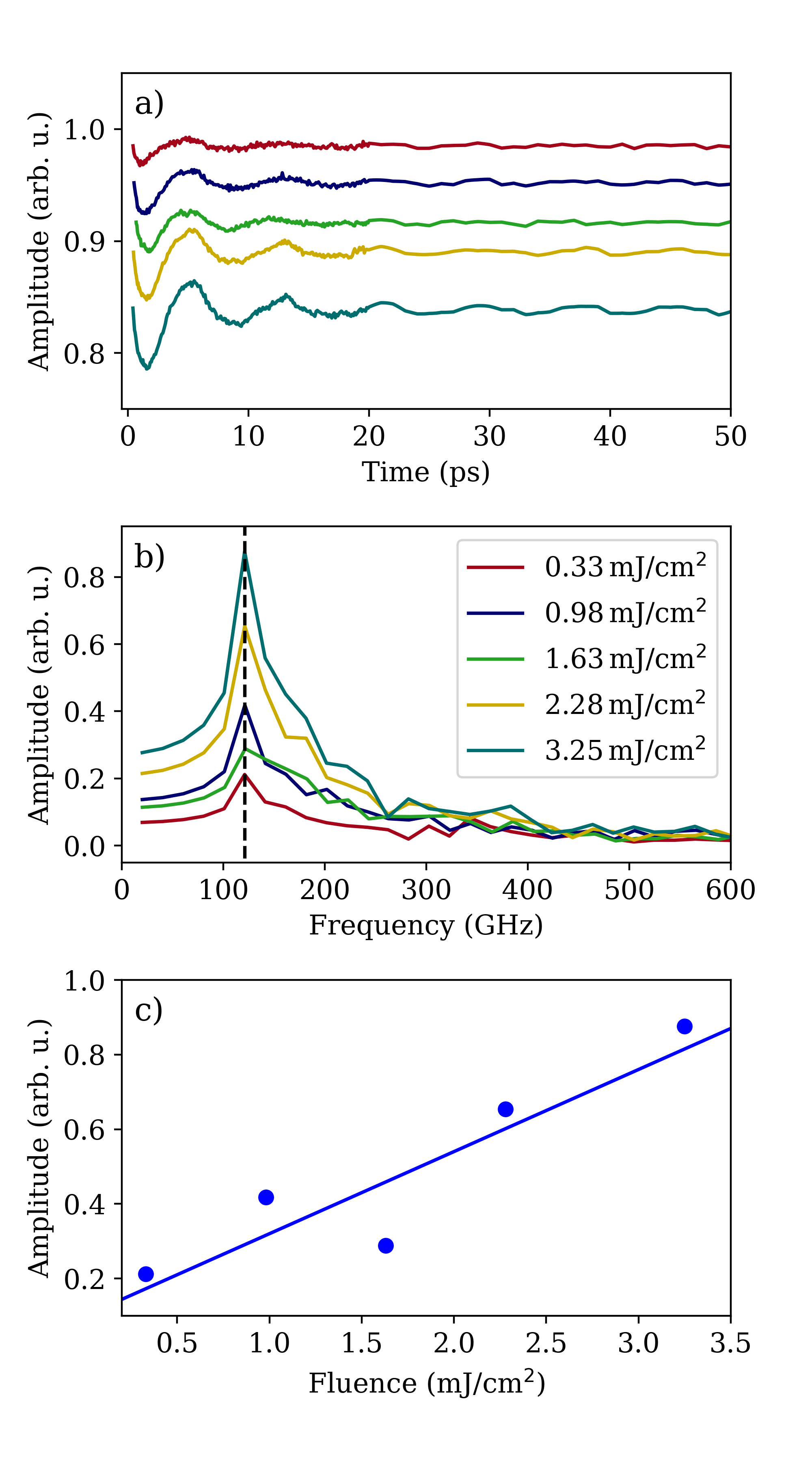}
    \caption{\textcolor{black}{(a) Extracted oscillatory component of the signals shown in Fig.~\ref{fig:mo_ptnio} after subtraction of the background associated with the reduction and relaxation of the AFM order parameter. (b) FFT spectra of the magnon oscillations. (c) Amplitude of the magnon oscillation as a function of pump fluence. The solid line is a guide to the eye.}}
    \label{fig:magnon}
\end{figure}

A discrepancy between the experimental and simulated MOBF traces in Fig.~\ref{fig:mo_ptnio} is observed in the lineshape near the minimum of the magneto-optical contrast, which becomes increasingly pronounced at higher pump fluences. To investigate the origin of this deviation, we extracted the oscillatory component of the signal by subtracting the background associated with the reduction and relaxation of the AFM order parameter, \textcolor{black}{leading to the curves} %{as} 
shown in Fig.~\ref{fig:magnon}(a). We subsequently performed a fast Fourier transform (FFT), with the resulting spectra presented in Fig.~\ref{fig:magnon}(b). 
For all pump fluences, a distinct peak is observed at approximately 130 GHz\textcolor{black}{. Its} 
FFT amplitude increases systematically with increasing fluence\textcolor{black}{, see Fig.~\ref{fig:magnon}(c).} 
This behavior is consistent with the excitation of the low-frequency magnon mode reported for thin NiO films in THz-emission and stimulated-Raman-scattering studies~\cite{qiu2021ultrafast, tzschaschel2017ultrafast, bossini2021ultrafast}.
The laser-induced excitation of this long-wavelength collective spin-wave mode is theoretically predicted: at the interface laser-generated hot electrons can launch long-range collective spin-wave excitations in AFMs on subpicosecond timescales \citep{ghosh2022driving}. 
The ultrafast, coherent laser-induced coupling between the excited electronic and spin subsystems can be understood in terms of transient, nonequilibrium modifications of the exchange interactions. 
These changes perturb the magnetic order and initiate collective spin-wave excitations, which can store part of the excess energy deposited in the electronic subsystem by the laser pulse and persist for several tens of picoseconds, see Fig.~\ref{fig:magnon}(a). 

Overall, our ultrafast MOBF experiments, in conjunction with TTM simulations, have provided compelling evidence of an indirect coupling of the optically excited electrons in Pt to the antiferromagnetic spin system of NiO, which is ultimately responsible for the ultrafast loss of antiferromagnetic order and for the generation of the low-frequency magnon mode in NiO.

\textcolor{black}{
Microscopically, there are two potential processes discussed in the literature for angular momentum transfer and dissipation at Pt/AFM material interfaces \cite{Moriyama2020, Kholid2023, Gray2019} 
that can be traced back to the specific properties of Pt. 
On the one hand, the strong atomic spin-orbit coupling of Pt can convert orbital and linear momentum of hot electrons into spin angular momentum, also into NiO by an interfacial exchange-mediated angular momentum transfer. 
Such ultrafast, direct coupling has also been found in a Pt/YIG bilayer structure~\cite{seifert2018femtosecond} and was interpreted as an interface-assisted spin transfer torque.
Due to the absence of magnetic order in Pt, however, the transferred angular momentum is expected to exhibit an average isotropic orientation
, thereby leading to an overall incoherent quenching of the antiferromagnetic order in NiO as also observed in our temperature-based simulations.}

\bigskip

\textcolor{black}{
In addition, the uncompensated magnetic moments of the interfacial Ni layer can induce a small static magnetization in Pt via the magnetic proximity effect, which follows the orientation of the interfacial Ni layer's magnetic moment in NiO, similar to other Pt/AFM or Pt/insulator systems \cite{Ma2025, Nakayama2013}.
Optical excitation of Pt can then in turn quench the proximity-effect-induced magnetic order, which causes an angular momentum dissipation from Pt into NiO again via interfacial exchange-mediated angular momentum transfer. 
This interfacial transfer is expected to lead to a spin-transfer torque due to the fixed spatial orientations of the proximity-induced magnetic moments in Pt, and thus can be responsible for generating a selected low-energy magnon mode in NiO. }
As there is no such strong magnetic proximity in the Al/NiO system, in comparison, these dynamics do not occur.

\section{Conclusion}

In conclusion, our joint experimental and theoretical study of the optically excited Pt/NiO bilayer system has uncovered an incoherent ultrafast loss of the antiferromagnetic order of NiO after optical excitation with fs pulses in the near-IR range, i.e., for photon energies below the NiO bandgap. The loss of the magnetic order occurs within the first ps after the optical excitation. 
Our experimental observation was attributed to an ultrafast energy transfer from the optically excited electron system in Pt directly into the antiferromagnetic spin system of NiO that depends on the coupling strength across the interface. 
This ultrafast energy transfer is accompanied by an interfacial, exchange-mediated transfer of angular momentum from the spin of Pt hot electrons to the spin of NiO.
The accumulation of spin angular momentum in Pt hot electrons is attributed to spin-orbit coupling in Pt and the magnetic proximity effect at the Pt/NiO interface.
Both interfacial coupling processes can be suppressed when either an insulating, nonmagnetic interlayer or a metal/AFM structure with a low spin-orbit coupling and lack of proximity effect is chosen such as Al.

Opening the perspective to steer both discussed mechanisms, our work lays the foundation for the ultrafast manipulation of the magnetic order of insulating and semiconducting (antiferromagnetic) materials by interface engineering and optical excitation below the optical bandgap.

\section*{Acknowledgements}

The  work was funded by the Deutsche Forschungsgemeinschaft (DFG, German Research Foundation) - TRR 173 - 268565370 Spin + X: spin in its collective environment (Projects A01, A03, A08, A11, and B03). 

B.S. further acknowledges funding by the Dynamics and Topology Research Center (TopDyn) funded by the State of Rhineland Palatinate.

R.R. acknowledges support from the Grants RYC 2019-026915-I and CNS2024-154679 funded by the MCIN/AEI/10.13039/501100011033 and by the ESF investing in your future and the European Union NextGenerationEU/PRTR,  the Xunta de Galicia (ED431F 2022/04, ED431C 2024/05, Centro Singular de Investigación do Sistema universitario de Galicia Accreditation 2023-2027, ED431G 2023/03) and the European Union (European Regional Development Fund - ERDF).

Additionally, T.K. and E.S. acknowledge support from JST-CREST (JPMJCR20C1 and JPMJCR20T2), Grant-in-Aid for Scientific Research (JP19H05600 and JP24K01326) from JSPS KAKENHI, Japan, and Institute for AI and Beyond of the University of Tokyo.

\newpage

\bibliographystyle{apsrev}
%\bibliography{bibfile/all} 

\end{document}